\documentstyle[eqsecnum,preprint,aps]{revtex}
\begin{document}
\draft
\preprint{CALT-68-1964}
\title{From Tetraquark to Hexaquark:
\\A Systematic Study of Heavy Exotics in the Large $N_c$ Limit}
\author{Chi-Keung Chow}
\address{California Institute of Technology, Pasadena, CA 91125}
\date{\today}
\maketitle
\begin{abstract}
A systematic study of multiquark exotics with one or $N_c-1$ heavy quarks
in the large $N_c$ limit is presented.
By binding a chiral soliton to a heavy meson, either a normal $N_c$-quark
baryon or an exotic $(N_c+2)$-quark baryon is obtained.
By replacing the heavy quark with $N_c-1$ heavy antiquarks, exotic
$(2N_c-2)$-quark and $2N_c$-quark mesons are obtained.
When $N_c = 3$, they are just the normal triquark baryon $Qqq$, the exotic
pentaquark baryon $Q\bar q\bar q\bar q\bar q$, tetraquark di-meson $\bar Q
\bar Q qq$ and the hexaquark di-baryon $\bar Q \bar Q \bar q \bar q \bar q
\bar q$ respectively.
Their stabilities and decays are also discussed.
In particular, it is shown that the ``heavy to heavy'' semileptonic decays
are described by the Isgur--Wise form factors of the normal baryons.
\end{abstract}

\narrowtext
\section{Introduction}
One of the early triumphs of the quark model is its success in describing the
hadron spectrum.
By regarding hadrons as $q\bar q$ or $qqq$ configurations, their quantum
numbers (charge, spin, isospin, strangeness, etc.) are well accounted for.
On the other hand, the question of the existence of exotics has attracted more
attention in recent years.
In particular, since Jaffe's classic papers \cite {j1,j2,j3} on di-meson and
di-baryon states in quark-bag model, the existences of multiquark states has
been investigated through many other approaches.

In this paper, we are going to investigate the existences of multiquark states
containing one or two heavy quarks.
In the large $N_c$ limit, they are states with one or $N_c-1$ heavy quarks.
Two types of interactions are going to play important roles in our discussion.
They are the heavy meson-chiral soliton binding, and the Coulumbic attraction
between heavy quarks.

By treating the low-lying baryons (N and $\Delta$) as chiral solitons
\cite{b0}, their interactions with heavy mesons have been previously calculated
in the large $N_c$ limit \cite {b1,b2,b3}.
It has been shown that there exist attractive channels between a heavy meson
and both a chiral soliton and an anti-soliton.
Hence we can systematically bind chiral solitons to heavy mesons, and study the
properties of those bound states with exotic quantum numbers.

On the other hand, in the heavy quark limit, the color potential between two
heavy quarks are Coulumbic.
As the result, $N_c-1$ heavy antiquarks can form a small colored complex of
size $(N_c\,\alpha_s(m_Q)m_Q)^{-1}$ which transforms just like a heavy quark
under color SU($N_c$).
(Note that we keep $N_c\,\alpha_s$ constant when taking the large $N_c$ limit;
the size of this ``fake heavy quark'' has a finite limit when
$N_c\rightarrow\infty$.)
Hence, for any hadron with one heavy quark, we can replace the heavy quark with
this ``fake heavy quark'' to obtain another hadron \cite {c1,c2}.
We will see below that some of these states are exotics.

In Section 2, we are going to review these two interactions by considering the
properties of normal baryons with one or $N_c-1$ heavy quarks.
Section 3 will investigate hadrons with $2N_c-2$, $N_c+2$ and $2N_c$ quarks,
which for brevity will be called tetraquarks, pentaquarks and hexaquarks
respectively.
Last of all, the validity of the results for finite quark masses and possible
generalizations of this framework is discussed in Section 4.

\section{Baryons with One or $N_c-1$ Heavy Quarks}
\subsection{Baryon with one heavy quark}
The simplest hadron with one heavy quark is the heavy meson $Q\bar q$.
The simplest light hadrons with non-zero baryon number are the low-lying
$q^{N_c}$ states, i.e., the nucleon N and the Delta $\Delta$, which can be
treated as chiral soliton in the large $N_c$ limit \cite{b0}.
Hence, in order to construct the simplest heavy baryon $Qq^{N_c-1}$, we
consider the binding between a heavy meson and a chiral soliton.

A classical chiral soliton is the configuration of the pion field satisfying
the hedgehog ansatz:
\begin{equation}
\Sigma_0=\exp(iF(x)\hat x\cdot\vec\tau),
\end{equation}
with $F(0)=-\pi$ and $F(\infty)=0$.
The interaction of a heavy meson and the pion field is given by the chiral
Lagrangian.
The binding potential between a heavy meson and a chiral soliton should
be expressible in terms of the heavy meson-pion coupling $g$ and the chiral
profile function $F(x)$ of the chiral soliton.
In Ref. \cite {b1,b2,b3}, it has been shown that this binding potential is
simple harmonic in the large $N_c$ limit and can be written in the form
\begin{equation}
V(x;K) = V_0(K) + \textstyle {1\over2} \kappa(K) x^2 ,
\end{equation}
where both the binding energy $V_0(K)$ and the spring constant $\kappa(K)$ are
odd functionals of $F(x)$.
The potential is independent of the heavy quark species (by heavy quark
symmetry) and the baryon spin (by the large $N_c$ limit), and solely depend on
$K=I+S_\ell$, where $I$ and $S_\ell$ are the isospin and spin of the light
degrees of freedom of the bound state, respectively.
For states with $K=0$, i.e, $(I,S_\ell)=(0,0)$, $(1,1)$ and {\it etc.},
$V_0=-\textstyle {3\over2} gF'(0)$ and the spring constant $\kappa$ is
positive.
Hence the resultant states are stable with the $(0,0)$ and $(1,1)$ states
identified as $\Lambda_Q$ and $\Sigma_Q$ respectively.
The $1/N_c$ corrections will break the degeneracy of these $K=0$ states with
the $(0,0)$ state as the ground state.
The value of $\kappa$ can be determined to be $(530 {\rm MeV})^3$ in the Skyrme
model and $(440 {\rm MeV})^3$ from $\Lambda^*_c-\Lambda_c$ splitting.
For states with $K=1$ the potential is opposite in sign:
\begin{equation}
V(x;K=1)=-\textstyle{1\over3}V(x;K=0),
\label{pr}
\end{equation}
and the resulting states are unbound.

Given the simple harmonic form of the binding potential, it is possible to
calculate the Isgur--Wise form factor $\eta(w)$, which describes the $Q_b
q^{N_c-1} \rightarrow Q_c q^{N_c-1}$ weak transition.
The Isgur--Wise form factor, which is the overlap of the light degrees of
freedom, is in this case just the overlap of simple harmonic wave functions
\cite {b3,b4}.
In the limit where $m_b$, $m_c\gg \Lambda_{\rm QCD}$, we have
\begin{equation}
\eta(w) = \int d^3 {\bf p}\, \phi^*({\bf p})\, \phi({\bf p} + {\bf k}) ,
\end{equation}
where
\begin{equation}
{\bf k} = m_S ({\bf v} - {\bf v'}),
\label{ho}
\end{equation}
and $\phi({\bf q})$ is the ground state simple harmonic wave function in the
momentum space.
In Eq. (\ref{ho}) $m_S$ is the mass of the chiral soliton.
The exact form of this integral is given in Ref. \cite{b3,b4}.
In particular we can verify the Luke's Theorem \cite{b5}, which states that
$\eta(w)$ is normalized to unity in the heavy quark limit.
\begin{equation}
\eta(1)=1,
\end{equation}
and the leading corrections are of order $(\Lambda_{\rm QCD}/m_Q)^2$.

The form factor $\eta(w)$ results from the non-perturbative interaction
between the chiral soliton and the heavy meson.
It describes all ``heavy to heavy'' transitions between normal baryons, like
$\Lambda_b \rightarrow \Lambda_c$ and $\Sigma_b \rightarrow \Sigma^{(*)}_c$.
Note that both transitions are described by the same form factor in the large
$N_c$ limit.  \cite {b4}

\subsection{Baryon with $N_c-1$ heavy antiquarks}
As discussed before, $N_c-1$ heavy antiquarks can form a small colored object
$\bar Q^{N_c-1}$ which transforms just like a heavy quark under color
SU($N_c$).
As long as $N_c\alpha_s(m_Q)m_Q\gg\Lambda_{\rm QCD}$, the light degrees of
freedom cannot resolve the individual heavy antiquarks within this ``fake heavy
quark.''
Hence, for any heavy hadron containing a single heavy quark, we can replace
this heavy quark with the ``fake heavy quark'' and obtain another hadron which
contains $N_c-1$ heavy antiquarks.
In particular, by replacing the heavy quark $Q$ in the heavy meson $Q \bar q$
with $\bar Q^{N_c-1}$, we get the non-exotic baryon  $\bar Q^{N_c-1} \bar q$.

$\bar Q^{N_c-1}$ has binding energy of the order of $N_c\,\alpha^2_s(m_Q)m_Q$.
When $m_Q \rightarrow \infty$, the binding energy grow to infinity.
The heavy antiquarks are very tightly bounded in the heavy quark limit,
and $\bar Q^{N_c-1} \bar q$ is safe from dissociations like $\bar Q^{N_c-1}
\bar q \rightarrow \bar Q q + \bar Q^{N_c-2} \bar q \bar q$.
As a result, $\bar Q^{N_c-1}$ must decay weakly.

To describe the weak decays, again we need an Isgur--Wise form factor.
For the decay $B_{a_1\dots a_{N_c-2} b} \rightarrow B_{a_1\dots a_{N_c-2} c}$,
where $B_{a_1\dots a_{N_c-2} b} \equiv \bar Q_{a_1}\dots \bar Q_{a_{N_c-2}}
\bar Q_b \bar q$ and $B_{a_1\dots a_{N_c-2} c} \equiv \bar Q_{a_1}\dots \bar
Q_{a_{N_c-2}} \bar Q_c \bar q$, the Isgur--Wise form factor $\eta_{a_1\dots
a_{N_c-2} (b\rightarrow c)} (w)$ is dominated by the overlap of the
Hartree--Fock wave functions, which describes the interaction of $\bar Q_b$ or
$\bar Q_c$ with the other $N_c-2$ heavy antiquarks \cite{c4}.
In the real world, $N_c=3$ and the Hartree--Fock wave functions become
Coulumbic wave functions \cite{c3}.
Hence we can calculate the Isgur--Wise form factor $\eta_{a(b\rightarrow c)}
(w)$ for the $\bar Q_a \bar Q_b \bar q \rightarrow \bar Q_a \bar Q_c \bar q$
transition by evaluating the overlap of Coulumbic wave functions.
(Note that $\eta_{a(b\rightarrow c)}(w)$ is called $\eta_{abc}(w)$ in Ref.
\cite{c3,t1}.)
For $B=\mu_{ab}\,\alpha_s(\mu_{ab})$ and $C=\mu_{ac}\,\alpha_s(\mu_{ac})$, we
have
\begin{equation}
\eta_{a(b\rightarrow c)}(w)=\int d^3{\bf p} \, \psi^* (C;{\bf p}) \, \psi
(B;{\bf p}+{\bf q}),
\end{equation}
where
\begin{equation}
{\bf q}=m_a({\bf v}-{\bf v'}),
\end{equation}
and $\psi (B;{\bf p})$ is the ground state Coulumbic wave function with Bohr
radius $B^{-1}$ and similar for $\psi (C;{\bf p+q})$.
The exact form of $\eta_{a(b\rightarrow c)} (w)$ is given in Ref. \cite{c3}.
Note that in general the Isgur--Wise form factor is not normalized at the point
of zero recoil.
In fact,
\begin{equation}
\eta_{a(b\rightarrow c)}(1)=\left({2\sqrt{BC}\over B+C}\right)^3 ,
\end{equation}
which is not equal to unity unless $B=C$, i.e, $m_b=m_c$.
This is very different from the normalization of $\eta(w)$, which holds
regardless of the size of $m_b-m_c$ as long as both $m_b$ and
$m_c\gg\Lambda_{\rm QCD}$.

The form factor $\eta_{a_1\dots a_{N_c-2} (b\rightarrow c)} (w)$ results from
the perturbative attraction between the heavy antiquarks.
As we will see below, to describe the semileptonic decays of the multiquark
exotics, we need just the two form factors we have discussed, namely $\eta(w)$
and $\eta_{a_1\dots a_{N_c-2} (b\rightarrow c)} (w)$.

\section{Heavy Exotics in the Large $N_{\lowercase{c}}$ Limit}
\subsection{Tetraquarks}
The large $N_c$ analog of tetraquark states $\bar Q \bar Qqq$ is ambiguous.
Both $\bar Q \bar Qqq$ and $\bar Q^{N_c-1} q^{N_c-1}$ reduce to $\bar Q \bar
Qqq$ when $N_c=3$.
In his classic paper on baryons in the $1/N_c$ expansion \cite{c4}, Witten
showed that $\bar Q \bar Qqq$ states are absent in the large $N_c$ limit.
On the other hand, stable $\bar Q^{N_c-1} q^{N_c-1}$ states, which he called
``baryonium'' states, exist.
In this section, we will discuss their properties and decay modes.

We can obtain a ``baryonium'' $\bar Q^{N_c-1}q^{N_c-1}$ by replacing the heavy
quark $Q$ in a heavy baryon $Qq^{N_c-1}$ with the ``fake heavy quark'' $\bar
Q^{N_c-1}$.
In the real world, where $N_c=3$, it is just the di-meson, or tetraquark hadron
$\bar Q \bar Qqq$ which Jaffe discussed in Ref. \cite {j1,j2} and subsequently
discussed in Ref. \cite{c4,t1,t2,t3,t4,t5,t6,t7}.

Since non-perturbative QCD cannot resolve the individual heavy antiquarks in
the ``fake heavy quark,'' the tetraquark has similar spectroscopic properties
with the normal baryon.
In particular, the lowest-lying configurations will have $(I,S_\ell)=(0,0)$.
These states are safe from dissociations like $\bar Q^{N_c-1}q^{N_c-1}
\rightarrow \bar Q^{N_c-1}\bar q + q^{N_c}$ and hence must decay weakly.

The Isgur--Wise form factor $\eta^{(4)}_{a_1\dots a_{N_c-2}(b\rightarrow c)}
(w)$ of the tetraquark decay $T_{a_1\dots a_{N_c-2}b} \rightarrow
T_{a_1\dots a_{N_c-2}c}$, where $T_{a_1\dots a_{N_c-2}b}\equiv\bar Q_{a_1}
\dots \bar Q_{a_{N_c-2}} \bar Q_bqq$ and $T_{a_1\dots a_{N_c-2}c}\equiv\bar
Q_{a_1}\dots Q_{a_{N_c-2}} \bar Q_cqq$ can be expressed as the product of two
terms with different physical origins \cite{t7}.
A perturbative term comes from the overlap of the initial and final
``fake heavy quark''  wave function.
Since this ``fake heavy quark'' transition is identical with what happens in
the $B_{a_1\dots a_{N_c-2} b} \rightarrow B_{a_1\dots a_{N_c-2} c}$ decay, the
perturbative term is exactly $\eta_{a_1\dots a_{N_c-2} (b\rightarrow c)}(w)$.
On the other hand, a non-perturbative term describes the overlap of the initial
and final light degrees of freedom under the color field of the ``fake heavy
quark.''
Since the light degrees of freedom of a tetraquark is identical to that of a
normal baryon, the non-perturbative term is just $\eta(w)$.
As a result, we get
\begin{equation}
\eta^{(4)}_{a_1\dots a_{N_c-2} (b\rightarrow c)}(w)
=\eta(w)\;\eta_{a_1\dots a_{N_c-2} (b\rightarrow c)}(w).
\end{equation}
When $N_c=3$ we have
\begin{equation}
\eta^{(4)}_{a(b\rightarrow c)}(w)=\eta(w)\;\eta_{a(b\rightarrow c)}(w),
\end{equation}
which has been proved in Ref. \cite{t7}.
Note that the normalization of $\eta^{(4)}_{a(b\rightarrow c)}(w)$ is given by
the normalization of each of its factors.
\begin{equation}
\eta^{(4)}_{a(b\rightarrow c)}(1)=\left({2\sqrt{BC}\over B+C}\right)^3.
\end{equation}
We will see this factorization of perturbative and non-perturbative
contributions also in the case of hexaquark decays.

\subsection{Pentaquarks}
In Ref. \cite{b2}, the authors has remarked upon the possibility of the
existence of exotic bound states of heavy mesons and chiral
{\it anti-}solitons.
In the large $N_c$ limit, a chiral anti-soliton, which has baryon number $-1$,
is also a configuration of the pion field satisfying the hedgehog ansatz.
The profile function satisfying the boundary condition $F(0)=\pi$ and
$F(\infty)=0$.
In other words, a chiral anti-soliton is obtained when we flip the sign of
$F(x)$ of a chiral soliton, and {\it vice versa}.
Since the binding potential $V(x)$ is odd in $F(x)$, the binding of a chiral
anti-soliton to a heavy meson will be the same in magnitude but opposite in
sign with that of a chiral soliton.
Denoting the binding of a chiral anti-soliton to a heavy meson by $\tilde
V(x;K)$, we have
\begin{equation}
\tilde V(x;K)=-V(x;K).
\label{pf}
\end{equation}
Hence $K=0$ states are unbound and the $K=1$ states are bounded.
The stable bound states are those with $(I,S_\ell)=(0,1)$, $(1,1)$,
$(1,0)$ and {\it etc.}
In the quark model, such states are exotic $Q\bar q^{N_c+1}$ multiquarks.
When $N_c=3$, these $Q\bar q\bar q\bar q\bar q$ states are just the pentaquarks
discussed in Ref. \cite{t4,p1,p2,p3,p4,p5}.

Combining Eq. (\ref{pr}) and Eq. (\ref{pf}), we get the relation
\begin{equation}
\tilde V(x;1)=\textstyle{1\over3}V(x;0),
\end{equation}
which means that the binding energy of the pentaquark is just a third of that
of a normal heavy baryon.
Still, the pentaquark is below the $Q\bar q^{N_c+1}\rightarrow Q\bar q + \bar
q^{N_c}$ threshold and hence must decay weakly.
Here again the Isgur--Wise function $\eta^{(5)}(w)$ is given by the overlap of
simple harmonic wave functions $\tilde\phi({\bf p})$.
The only difference is that for the pentaquark system the spring constant is
just $\textstyle{1\over3}\kappa$, i.e., a third of that of a normal heavy
baryon.
Since the natural unit for momentum of a simple harmonic oscillator is
$(m_S\kappa)^{1/4}$, we have
\begin{equation}
\tilde\phi({\bf p})=\phi(3^{1/4}{\bf p}).
\end{equation}
With
\begin{equation}
\eta^{(5)}(w) = \int d^3 {\bf p}\, \tilde\phi^*({\bf p})\,
\tilde\phi({\bf p} + {\bf k}) ,
\end{equation}
we finally obtain
\begin{equation}
\eta^{(5)}(w)=\eta(\sqrt{3}(w-1)+1).
\end{equation}

We have succeeded in relating the Isgur--Wise form factors of pentaquarks and
that of normal heavy baryons.
Like $\eta(w)$, $\eta^{(5)}(w)$ also obeys Luke's theorem and is normalized at
the point of zero recoil,
\begin{equation}
\eta^{(5)}(1)=1.
\end{equation}

We can also consider the hadron obtained by replacing the heavy quark in a
pentaquark system by a ``fake heavy quark.''
The resultant hadron is the famous $H$-dibaryon, which is also known as the
hexaquark.

\subsection{Hexaquarks}
As suggested above, when we replace the heavy quark inside the pentaquark
system with a ``fake heavy quark,'' the resultant system $\bar Q^{N_c-1} \bar q
^{N_c+1}$ has baryon number $2$.
When $N_c=3$, the hadron $\bar Q \bar Q \bar q \bar q \bar q \bar q$ is just
the $H$ particle first suggested by Jaffe \cite{b3} and subsequently discussed
by Ref. \cite{c4,t4,p3,h1,h2,h3,h4,h5,h6,h7,h8,h9,h10,h11}.
It is one of the most well-discussed exotics as it arises in many different
scenarios like the bag models, large $N_c$ Hartree-Fock model, $\Lambda\Lambda$
molecule, Skyrme models, potential models and lattice gauge calculations.
Noteworthy are Ref. \cite{h3} and Ref. \cite{h7}, in which $H$ arises as
topological chiral solitons under SU(3)$_L \times$ SU(3)$_R$ chiral symmetry.
While the normal chiral solitons (N and $\Delta$) are pion configurations in
the SU(2) submanifold spanned by the SU(3) generators $\{ \lambda_1, \lambda_2,
\lambda_3 \}$, $H$ are pion configurations in the SU(2) submanifold spanned by
$\{ \lambda_2, \lambda_5, \lambda_7 \}$.
Due to this relation between the structure constants of SU(3),
\begin{equation}
f_{257} = \textstyle{1\over2} f_{123},
\end{equation}
the baryon number of $H$ is twice that of a normal chiral soliton.
This SU(3) group theoretical approach is applicable when there are three light
flavors.
It is interesting to see that, in our formalism, hexaquarks also arise through
the interplay of SU(2)$_L\times$ SU(2)$_R$ chiral symmetry and heavy quark
symmetry.

Since the light degrees of freedom of a hexaquark is identical to that of a
pentaquark, we have $K=1$ stable configurations.
The discussion on the stability of the hexaquark runs parallel to that of the
tetraquark.
The large chromoelectric binding energy prevent the dissociation $\bar
Q^{N_c-1} \bar q^{N_c+1}\rightarrow \bar Q \bar q^{N_c-1} + \bar Q^{N_c-2} \bar
q \bar q$ or other decay modes which involves the splitting up of the ``fake
heavy quark'' system.
On the other hand, the stability of the pentaquark system guarantees the
stabililty of the hexaquark against $\bar Q^{N_c-1} \bar q^{N_c+1} \rightarrow
\bar Q^{N_c-1} \bar q + \bar q^{N_c}$.
Hence the hexaquark is stable with respect to strong interactions and decays
weakly.

Just like its tetraquark counterpart, the hexaquark Isgur--Wise weak form
factor $\eta^{(6)}_{a_1\dots a_{N_c-2}(b\rightarrow c)}
(w)$ of the hexaquark decay $H_{a_1\dots a_{N_c-2}b} \rightarrow
H_{a_1\dots a_{N_c-2}c}$, where $H_{a_1\dots a_{N_c-2}b}\equiv\bar Q_{a_1}
\dots \bar Q_{a_{N_c-2}} \bar Q_b \bar q^{N_c+1}$ and $H_{a_1\dots
a_{N_c-2}c}\equiv\bar Q_{a_1}\dots Q_{a_{N_c-2}} \bar Q_c \bar q^{N_c+1}$ can
also be expressed as the product of a perturbative factor and a
non-perturbative factor.
The perturbative part, which results from the chromoelectric attraction between
the heavy antiquarks, is again given by $\eta_{a_1\dots a_{N_c-2} (b\rightarrow
c)}(w)$.
The non-perturbative part, which describes the overlap between the initial and
final light degrees of freedom, is given by $\eta^{(5)}(w)$ as the light
degrees of freedom of a hexaquark is identical with that of a pentaquark.
As a result,
\begin{equation}
\eta^{(6)}_{a_1\dots a_{N_c-2}(b\rightarrow c)}(w) =
\eta^{(5)}(w)\; \eta_{a_1\dots a_{N_c-2} (b\rightarrow c)}(w).
\end{equation}
When $N_c=3$ we have
\begin{equation}
\eta^{(6)}_{a(b\rightarrow c)}(w)=\eta^{(5)}(w)\;\eta_{a(b\rightarrow c)}(w).
\end{equation}
Last of all, we again have the normalization condition
\begin{equation}
\eta^{(6)}_{a(b\rightarrow c)}(1)=\left({2\sqrt{BC}\over B+C}\right)^3.
\end{equation}

\section{Conclusion}
In summary, the following results have been obtained.

1)  In the large $N_c$ limits, there exists heavy tetraquark, pentaquark, and
hexaquark exotic states which are stable with respect to strong interactions.
It is noted that the stabilities of tetraquarks depends simply on the heavy
quark limit and the stabilities of normal $\Lambda_Q$ baryons, while the
stabilities of pentaquarks and hexaquarks have been shown in the context of the
chiral soliton model, which is a crucial assumption in our discussion.

2)  The $(I, s_\ell)$ of these multiquark states are known.

3)  The ``heavy to heavy'' weak decay of each of these multiquark states are
described by a single Isgur--Wise form factor.  Moreover, these Isgur--Wise
form factors can be expressed in terms of the Isgur--Wise form factors of
normal baryons.

Naturally the question of whether in this framework one can generate higher
multiquark states like heptaquarks and octaquarks arises.
To answer this question, note that all the hadrons considered in this paper can
be reduced to two-body bound states, of either a heavy meson or ``fake heavy
meson'' in one hand, and a chiral soliton or anti-soliton in the other.
It is exactly the possibility of such reduction to a two-body problem which
simplifies the systems and enable us to get the results listed above.
When we move on higher multiquark states, such reductions are impossible.
For example, the heptaquark is a ``fake heavy meson'' $\bar Q \bar Q \bar q$
bounded to {\it two} chiral solitons $qqq$, and the octaquark is a heavy meson
$Q \bar q$ bounded to {\it two} chiral anti-solitons $\bar q \bar q \bar q$.
In general, such three-body systems are intractable.
Hence we do not expect a simple generalization of our framework to describe
these higher multiquark states.

Our discussion on hadrons with $N_c-1$ heavy antiquarks depends crucially on
the assumption that the light degrees of freedom cannot resolve the ``fake
heavy quark'' or equivalently $N_c\alpha_s(m_Q)m_Q\gg\Lambda_{\rm QCD}$.
In the real world, since the top quark does not live long enough to form
hadrons, we just have two ``hadronizable'' heavy quarks, the $b$-quark and the
$c$-quark.
The assumption above, however, holds for neither of them, and our results
cannot be applied directly.
Still, it is possible that the picture above is at least qualitatively correct
and can serve as the starting point of quantitative investigations of the heavy
multiquark systems by including the effects of $1/m_Q$ corrections.

\bigskip
This work was supported in part by the U.S. Dept. of Energy under Grant No.
DE-FG03-92-ER 40701.

\end{document}